\documentclass[sn-chicago]{sn-jnl}

\usepackage{graphicx}%
\usepackage{multirow}%
\usepackage{amsmath,amssymb,amsfonts}%
\usepackage{amsthm}%
\usepackage{mathrsfs}%
\usepackage[title]{appendix}%
\usepackage{xcolor}%
\usepackage{textcomp}%
\usepackage{manyfoot}%
\usepackage{booktabs}%
\usepackage{algorithm}%
\usepackage{algorithmicx}%
\usepackage{algpseudocode}%
\usepackage{listings}%
\usepackage{multicol}

\theoremstyle{thmstyleone}%
\newtheorem{theorem}{Theorem}
\newtheorem{proposition}[theorem]{Proposition}%
\newtheorem{corollary}[theorem]{Corollary}

\theoremstyle{thmstyletwo}%
\newtheorem{remark}{Remark}%

\theoremstyle{thmstylethree}%

\usepackage{dsfont}
\usepackage{orcidlink}
\usepackage{float}	
\usepackage{mathbbol}
\usepackage[justification=centering]{caption}
\usepackage{enumitem}
\usepackage{array,multirow}
\usepackage{caption}
\usepackage{hyperref}
\hypersetup{
	colorlinks=true,
	linkcolor=blue,
	filecolor=magenta,      
	urlcolor=blue,
	citecolor=blue,
}
\renewcommand{\theequation}{\thesection.\arabic{equation}}

\newcommand{\Newline}{\phantom{x} \\ \noindent}

\newcommand{\textbi}[1]{\textbf{\textit{#1}}}

\newcommand{\m}{\mbox{}}

\newcommand{\vertspace}{\vspace{6pt}}

\newcommand{\epsr}{\varepsilon_{\text{rel}}}
\newcommand{\Sa}{S_{\text{pre}}}
\newcommand{\Sb}{S_{\text{post}}}
\newcommand{\Spre}{\overline{S}_{\text{pre}}}
\newcommand{\Spo}{\overline{S}_{\text{post}}}

\newcommand{\BN}{\mathcal{BN}}
\newcommand{\N}{\mathcal{N}}

\newcommand{\s}{\sigma}
\renewcommand{\m}{\mu}
\newcommand{\spo}{\sigma_{\text{post}}}
\newcommand{\spre}{\sigma_{\text{pre}}}
\newcommand{\mpo}{\m_{\text{post}}}
\newcommand{\mpre}{\m_{\text{pre}}}
\newcommand{\rba}{\rho_{\text{pre,post}}}
\newcommand{\rs}{\rho^{*}}
\newcommand{\pr}{\operatorname{Pr}}

\newcommand{\Hg}{\hat{g}}

\newcommand{\nvertspace}{\vspace{-6pt}}

\newcommand{\E}[1]{E\left\{#1\right\}}

\newcommand{\Di}[1]{D\left\{#1\right\}}

\begin{document}

\title[Novel computational approaches for ratio distributions]{Novel computational approaches for ratio distributions with an application to Hake's ratio in effect size measurement}

\author*[1]{\fnm{Jozef} \sur{Han\v{c} \orcidlink{0000-0003-1359-6117}}}\email{jozef.hanc@upjs.sk}

\author[2]{\fnm{Martina} \sur{Han\v{c}ov\'a} \orcidlink{0000-0001-8004-3972}}\email{martina.hancova@upjs.sk}

\author[1]{\fnm{Dominik} \sur{Borovsk\'y}}\email{dominik.borovsky@student.upjs.sk}

\affil*[1]{\orgdiv{Institute of Physics}, \orgname{P.J. \v{S}af\'arik University}, \orgaddress{\street{Park Angelinum 9}, \city{Ko\v{s}ice}, \postcode{04001}, \country{Slovakia}}}

\affil[2]{\orgdiv{Institute of Mathematics}, \orgname{P.J. \v{S}af\'arik University}, \orgaddress{\street{Jesenn\'a 5}, \city{Ko\v{s}ice}, \postcode{04001}, \country{Slovakia}}}

\abstract{
	
Ratio statistics and distributions are fundamental in various disciplines, including linear regression, metrology, nuclear physics, operations research, econometrics, biostatistics, genetics, and engineering.
In this work, we introduce two novel computational approaches for evaluating ratio distributions using open data science tools and modern numerical quadratures. 
The first approach employs 1D double exponential quadrature of the Mellin convolution with/without barycentric interpolation, which is a very fast and efficient quadrature technique. The second approach utilizes 2D vectorized Broda-Khan numerical inversion of characteristic functions. It offers broader applicability by not requiring knowledge of PDFs or the independence of ratio constituents. 
The pilot numerical study, conducted in the context of Hake’s ratio — a widely used measure of effect size and educational effectiveness in physics education — demonstrates the proposed methods' speed, accuracy, and reliability. The analytical and numerical explorations also provide more clarifying insight into the theoretical and empirical properties of Hake’s ratio distribution. The proposed methods appear promising in a robust framework for fast and exact ratio distribution computations beyond normal random variables, with potential applications in multidimensional statistics and uncertainty analysis in metrology, where precise and reliable data handling is essential.}
		



\maketitle
\newpage

\section{Introduction}\label{sec1}

Today, ratio statistics and their distributions play an essential role not only in linear regression theory and metrology but also in a broad range of applications. According to several works \citep{marsaglia_ratios_2006, nadarajah_note_2006, nadarajah_linear_2011, celano_statistical_2014, perri_environmental_2021}, examples include nuclear physics (mass-to-energy ratios), operations research and engineering (safety factors in design, signal-to-noise ratios, radar distributions), econometrics (economic indicators), biostatistics (enzyme activity, red blood cell lifespan, medical study ratios), genetics (Mendelian inheritance ratios), and the food and pharmaceutical industries (digestibility measures, component ratios in foods or drugs), as well as meteorology (target-to-control precipitation ratios) and environmental science (environmental concentrations and loads). 

If we consider the ratio of normal random variables (RVs), \cite{diaz-frances_existence_2013} point to the need to examine and apply this distribution in fields such as cytometry, physiology, risk analysis, and DNA microarrays. In this paper, we will present another application — ratios used in social, behavioral, and educational sciences as statistical measures of effect size (ES) for assessing intervention improvements, either pre- and post-intervention or between control and experimental groups.

However, the analytical complexity of ratio statistics and distributions often necessitates efficient computational techniques and software for fast and accurate estimation and inference. Today, in the era of open data science, where users of R and Python expect a package for nearly every statistical application, it is surprising that, to our knowledge, no specialized package exists for the ratio of normal random variables to compute the probability density function (PDF), cumulative distribution function (CDF), approximations, or perform statistical inference. The only available software remains a 20-year-old C code provided as supplementary material of \cite{marsaglia_ratios_2006}. 

As we describe in detail later, the more general Python package PACAL \citep{korzen_pacal_2014} is capable of numerically performing arithmetic operations ($+, -, *, /$) on RVs that have a known analytic form of their PDFs. If we consider existing approaches based on the characteristic function (CF) of RVs, numerical inversion of the CF into a PDF (CDF) for the ratio of RVs has been established only for non-negative RVs via logarithmic transformation, assuming the CF of the corresponding log-transformed RVs is known \citep{witkovsky_numerical_2016, witkovsky_charfuntool_2023}.

Problematic computational issues may even arise when using known analytic formulas in computer algebra systems, such as those involving confluent hypergeometric  functions \citep{johansson_computing_2019,hancova_practical_2022}.In metrology, the official guide on uncertainty in measurements officially recommends a purely simulation-based MC approach \citep{korzen_pacal_2014, witkovsky_numerical_2016}, but its main weakness is very slow convergence and low accuracy.

Based on these considerations, our goal is to explore both existing and potential numerical approaches and corresponding digital tools that are suitable for computing ratio distributions. The goal appears particularly important in the context of computational research, which is emerging as the third paradigm in mathematics and science. Within this paradigm, complex and systematic numerical studies and simulations also play a crucial role in statistics, proving indispensable for refining, validating, and advancing theoretical developments and practical applications.

Regarding the structure, the paper is divided into two main parts. In the first part (Sec. 2), we focus on the ratio of correlated normal random variables, with a practical illustration of Hake's ratio for effect size measurement. This section reviews the most relevant statistical properties of such a ratio, covering analytical forms, graphical properties, and numerical approximations. They serve as an essential foundation for pilot testing in the second part of the paper (Sec. 3 \& 4).

In Sec. 3, we address recent computational and numerical approaches for ratio distribution calculations, mainly connected to integral transforms (i.e., Mellin, Fourier), including our two novel computational approaches that leverage open data science tools and modern numerical quadratures. In Sec. 4, all these approaches are compared in a pilot numerical study on calculating the ratio of independent normal RVs, in the context of Hake's ratio,  where we apply analytical and graphical tools from Sec. 2 as a necessary cross-check of the accuracy, reliability  and stability of the given methods.  

In conclusions, we analyze the significance of our statistical findings and discuss their broader implications and potential future applications. Particularly, the preliminary results of the proposed computational methods indicate that these methods can be applied beyond the specific ratio of normal variables, enabling rapid data analysis based on exact probability distributions for ratios of non-normal (un)correlated random variables. This capability holds promise for applications in multidimensional statistics and uncertainty analysis in metrology, where precise data handling is crucial. For clarity, the interpretation of Hake's ratio and detailed proofs, formulas, algorithms, and computational tools used are in the appendices.

\section{Basic statistical properties of Hake's ratio}\label{sec:2 properties}

In social and behavioral sciences, medicine, and education, research and statistical analysis dealing with subjective repeated measurements often require ES reporting \citep{grissom_effect_2012, bowman_importance_2017, cumming_introduction_2024}. 

For this purpose, in physics education and physics education research (PER), a widely used measure is a ratio, introduced by R. Hake \citep{hake_interactive-engagement_1998}, also known as the average normalized gain \citep{nissen_comparison_2018, mckagan_physport_2020}. Hake's statistics $g$ and its sample-based estimator $\hat{g}$ can be defined as
\begin{equation}
	g = \frac{\mpo - \mpre}{100 - \mpre}, \qquad \Hg = \frac{\Spo - \Spre}{100 - \Spre},
	\label{eq:normalized_gain}
\end{equation}
\noindent where $\Spre$ and $\Spo$ are the average relative scores (in \%) of a given group of $n$ participants on a diagnostic measurement tool, such as a standardized didactic test; $\mpre,\mpo$ represent the population mean scores $\E{\Sa},\E{\Sb}$.

\subsection{Analytic forms}

In educational settings, as stated in Prop.~\ref{prop:g_distribution} in Appendix \ref{app:proofs}, Hake's ratio $\Hg$ appears to be a ratio of two correlated normal RVs. Therefore, we briefly summarize the relevant results of the statistical literature regarding the ratio's analytical properties.

In \citet[p.~636]{hinkley_ratio_1969}, we can find the analytic forms for the PDF and CDF. The following theorem summarizes the PDF result. 

\vertspace
\begin{theorem}[PDF of the ratio — analytic form]\label{thm:hinkley_pdf_W} \Newline
	Let $W = X_1 / X_2$, $(X_1, X_2) \sim \mathcal{BN}(\mu_1, \mu_2; \s_1, \s_2; \rho)$. Then the PDF of $W$ is given by 
	\nvertspace \begin{equation}
		\begin{aligned}
			f_W(w) = & \, \frac{b(w) d(w)}{\sqrt{2 \pi} \sigma_1 \sigma_2 a^3(w)} 
			\left[\Phi\left(\frac{b(w)}{\sqrt{\left(1-\rho^2\right) a(w)}}\right) - \Phi\left(-\frac{b(w)}{\sqrt{\left(1-\rho^2\right) a(w)}}\right)\right] \\
			& + \frac{\sqrt{1-\rho^2}}{\pi \sigma_1 \sigma_2 a^2(w)} \exp \left(-\frac{c}{2\left(1-\rho^2\right)}\right),
		\end{aligned}
	\end{equation}
	where $\Phi$ is the CDF of $\mathcal{N}(0,1)$ \\[-18pt]
	\begin{align*}
		a(w) &= \left(\frac{w^2}{\sigma_1^2} - \frac{2 \rho w}{\sigma_1 \sigma_2} + \frac{1}{\sigma_2^2}\right)^{1/2}, \quad
		b(w) = \frac{\mu_1 w}{\sigma_1^2} - \frac{\rho (\mu_1 + \mu_2 w)}{\sigma_1 \sigma_2} + \frac{\mu_2}{\sigma_2^2}, \\
		c &= \frac{\mu_1^2}{\sigma_1^2} - 2 \rho \frac{\mu_1 \mu_2}{\sigma_1 \sigma_2} + \frac{\mu_2^2}{\sigma_2^2}, \quad
		d(w) = \exp \left(\frac{b^2(w) - c a^2(w)}{2 (1 - \rho^2) a^2(w)}\right).
	\end{align*}
\end{theorem}

\noindent The symbol $\BN$ denotes a bivariate normal distribution with given parameters.

However, a more comprehensible analytical form also exists, which is both clearer for theoretical analysis and more suitable for practical computational implementation. Based on the works of \citet[Prop. on p.~2]{marsaglia_ratios_2006} and \citet[Prop.~3 and Lem. 4, p.~1587]{pham-gia_density_2007}, we can formulate the following theorem.

\vertspace

\begin{theorem}[PDF of the ratio —  alternative analytic form]\label{thm:pham-gia_pdf_T}\Newline
	For $W = X_1 / X_2$, where  $(X_1, X_2) \sim \mathcal{BN}(\mu_1, \mu_2; \s_1, \s_2; \rho)$, there exist real constants $r$ and $s$ such that 
	\vertspace
	\begin{itemize}
		\item[] $r\left(W - s \right)$ is distributed as $T = \dfrac{a + V_1}{b + V_2}$, 
			\item[] $(V_1, V_2) \sim \BN(0, 0; 1, 1; 0)$, {\small $b =  \dfrac{|\m_2|}{\s_2}$, $a =  \dfrac{\m_1 / \s_1 - \rho \m_2 / \s_2}{\pm\sqrt{1 - \rho^2}} \geq 0$} 
		\nvertspace \begin{equation}
			f_T(t) = \frac{1}{\exp\left(\frac{a^2 + b^2}{2}\right) \pi (1 + t^2)} \, {}_1F_1\left( \begin{matrix} \tiny {1} \\ {1/2} \end{matrix}\,; \frac{(at + b)^2}{2(1 + t^2)}\right),
			\label{eq:f_T_pham-gia}
		\end{equation}
		\item[] $W$ is distributed as $\dfrac{1}{r}T + s$, where $s=\tfrac{\s_1}{\s_2}\rho, 1/r=\pm\tfrac{\s_1}{\s_2} \sqrt{1 - \rho^2} $
		\nvertspace \begin{equation}
			f_W(w) = \frac{1}{\pm\frac{\s_1}{\s_2} \sqrt{1 - \rho^2}} f_T\left(\frac{w - \rho \frac{\s_1}{\s_2}}{\pm\frac{\s_1}{\s_2} \sqrt{1 - \rho^2}}\right),\label{eq:f_W_pham-gia}
		\end{equation}
	
\item[] where ${}_1 F_1\left(\begin{matrix} \alpha \\ \beta \end{matrix} ; z\right)$ is Kummer's confluent hypergeometric function.
\end{itemize}
\end{theorem}

\vertspace
The sign of the denominator in $a$ for \eqref{eq:f_T_pham-gia}  is chosen such that $a \geq 0$ always holds; the same sign is subsequently applied to $r$ and $f_W$ in \eqref{eq:f_W_pham-gia}.

\noindent A surprising result of the theorem, particularly for non-statisticians, is that one need not study the ratio $W$ of correlated normal RVs. It suffices to analyze the transformed ratio $T$ of independent normal RVs  with nonnegative means $a$ and $b$.

The Kummer confluent hypergeometric function \citep{gradshteyn_table_2007} with given parameters can be expressed via the error function \citep{hancova_practical_2022}
\nvertspace
\begin{equation}
	{}_1F_1\left( \begin{matrix} \tiny {1} \\ {1/2} \end{matrix}\,; \dfrac{q^2}{2}\right) = \left( \sqrt{\dfrac{\pi}{2}} q\right)  \operatorname{erf} \left(\dfrac{q}{\sqrt{2}}\right)\exp{\left( \dfrac{q^2}{2}\right)}+1, \quad q=\dfrac{b+a t}{\sqrt{1+t^2}}.	\label{eq:FvsERF}
\end{equation}
In the context of possible computational issues with confluent hypergeometric functions mentioned in the introduction, this form becomes a key component of computational implementation, as the standard computation of the error function $\operatorname{erf}$ is typically much faster and more reliable. We utilize it in our pilot numerical study (Sec. 4) to generate PDF values with quadruple precision, a choice made specifically for error control and stability assessment.

\vertspace
\begin{remark} \small
	It appears that in the statistical literature, the most cited source is \cite{hinkley_ratio_1969}, even though it does not derive the analytical form but compares the exact distribution with its approximation. The derivation itself was already provided earlier in \cite{fieller_distribution_1932}; therefore, in some resources, the ratio of normal RVs is referred to as Fieller's distribution.
It is also worth noting that the idea of using $T$ as sufficient for studying the properties of $W$ was already suggested by \cite{marsaglia_ratios_1965}, prior to \cite{hinkley_ratio_1969}. However, this connection was not fully recognized by \cite{hinkley_ratio_1969} and was later clarified in \cite{hinkley_correction_1970}.
\end{remark}

\subsection{Shapes and approximations}

While the previous section provided analytical tools for our numerical study, this section focuses on graphical properties, which can serve as qualitative cross-checking tools for the reliability of numerical simulations, such as the automatic detection of the modality (number of peaks) of a numerically generated distribution.

The formally simpler expression \eqref{eq:f_T_pham-gia} for PDFs in terms of ${}_1F_1$ allows for examining properties such as symmetry, modality, and monotonicity, as discussed in \cite{pham-gia_density_2007}. Concerning modality, mathematically described by the number of solutions to $f_T(t)' = 0$, we can obtain an empirical rule for any $b$ \citep{marsaglia_ratios_2006}

\vertspace
\begin{center}
	\textit{Rule of Thumb:} For $a \le 1$ the PDF of $T$ (and $W$) is \textit{unimodal},\\ \phantom{Rule of Thumb: } for $a \ge 2.256$ the PDF $T$ (and $W$) is \textit{bimodal}.
\end{center}
\vertspace

\noindent In the "transition range" $1 \le a \le 2.256$, the modality depends on $b$ and is determined by the modality curve (see Appendix \ref{app:proofs}). 

Fig.~\ref{fig:bimodal_distribution} illustrates the bimodal shape of Hake's ratio $f_W$ for specific values of the original parameters, resulting in $(a, b) = (2, 0.25)$. Here, we used formulas for $a$ and $b$ from Cor.~\ref{cor:Hake_ab} (Appendix \ref{app:proofs}), an immediate consequence of  Prop.~\ref{prop:g_distribution} and Thm.~\ref{thm:pham-gia_pdf_T}.

\begin{figure}[h!]
	\centering
	\includegraphics[width=\textwidth]{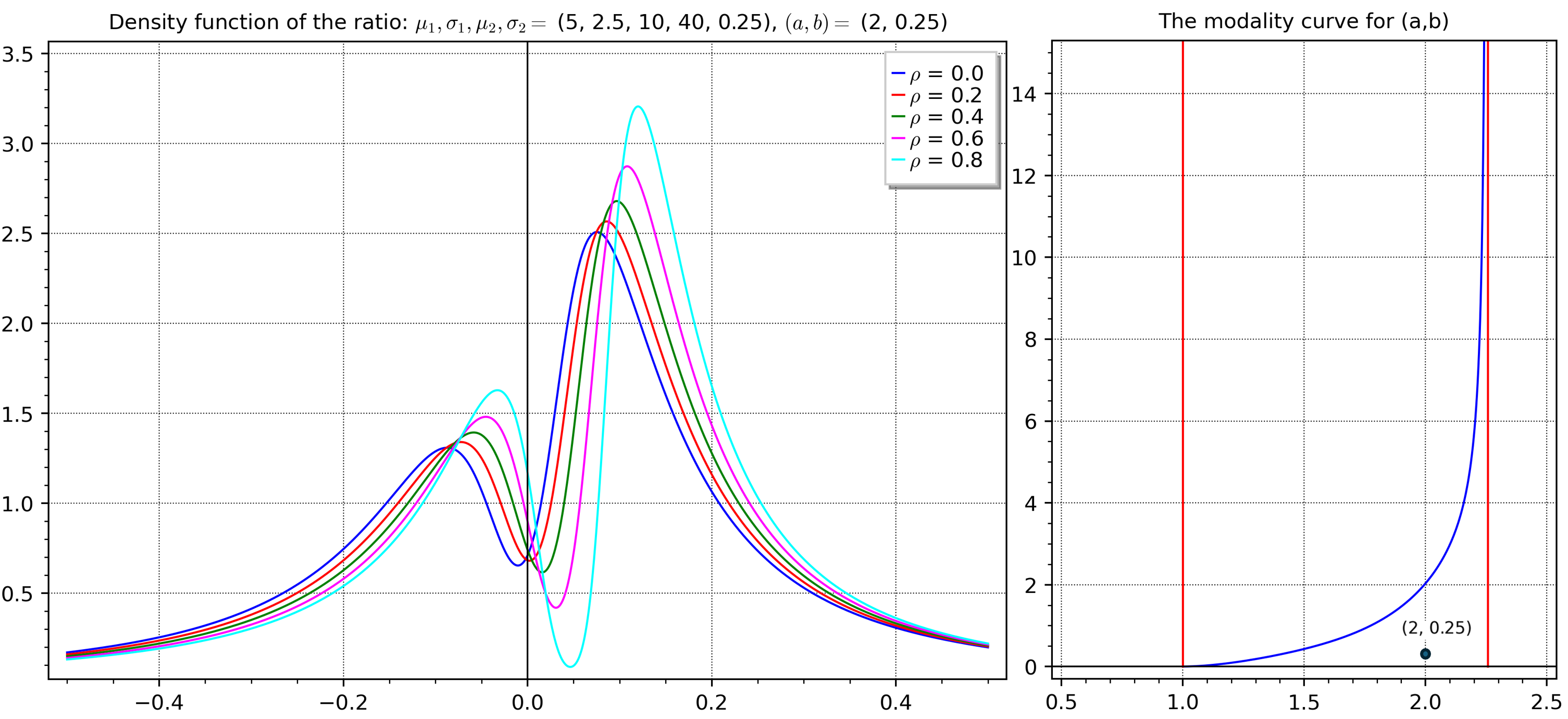}
	\caption{The shape of $W$ for a set of parameter values illustrating bimodality.}
	\label{fig:bimodal_distribution}
\end{figure}

\vertspace
\begin{remark} \small
	Sec. 5 in \cite{pham-gia_density_2007} provides a rigorous procedure for dividing the $(a, b)$ plane into two regions by a modality curve given by $f_T(t)' = 0$, such that the PDF is unimodal for points on the left of that curve and bimodal on its right (our Fig.~\ref{fig:bimodal_distribution}, right). Modern digital tools \citep{gajdos_interactive_2022} allow the straightforward creation of an interactive Jupyter notebook demonstrating the entire process as a virtual experiment in a highly visual and intuitive way.
\end{remark}

\vertspace

Another surprising and statistically inconvenient property of the distribution of the ratio of two normal random variables, $X_1$ and $X_2$, is its heavy-tailed nature and lack of finite moments. However, in the case of a unimodal shape, empirical data of Hake's ratio \citep{coletta_evidence_2023} typically result in a distribution that can be closely approximated by a normal distribution. This behavior of Hake's ratio is also explainable by previously published statistical works. Indeed, mathematical conditions for such a reasonable approximation can be formulated numerically \citep{marsaglia_ratios_2006} and through a rigorous existence theorem \citep{diaz-frances_existence_2013}.

According to \cite{diaz-frances_existence_2013}, if we consider independent normal RVs, $X_1 \sim \N(\mu_1, \sigma_1^2)$ and $X_2 \sim \N(\mu_2, \sigma_2^2)$, with strictly positive means and variances, and coefficients of variation $\delta_1= \sigma_1/\mu_1$, $\delta_2= \sigma_2/\mu_2$ less than 1, then by applying a Taylor series expansion of the ratio $X_1/X_2$ around the point $(\mu_1, \mu_2)$, the first-order term provides a mean, while the second-order term provides the variance of the normal distribution approximation $\N(\mu, \sigma^2)$ with the following CDF
\nvertspace
\begin{equation} G\left(z ; \mu, \delta_1, \delta_2\right)=\Phi\left(\frac{z-\mu}{\mu \sqrt{\delta_1^2+\delta_2^2}}\right),\label{eq:diaz-francesNappr} \\[-6pt]
 \end{equation} 
\noindent where $\mu = \mu_1/\mu_2, \sigma^2 = \mu^2 (\delta_1^2+\delta_2^2)$ and $\delta^2 =\sigma^2/ \mu^2 = \delta_1^2+\delta_2^2$. These "pseudo-moments", called by \cite{marsaglia_ratios_2006} practical moments, also correspond to the well-known law of uncertainty propagation known from metrology. In our pilot study, we use these moments for a raw estimation of an interval where we perform and test numerical calculations for the PDF (more specific details are in Sec.~4.1).

 \cite{diaz-frances_existence_2013} also introduce another graphical tool, the ROC curve, which is useful for assessing the closeness of the distribution of a specific ratio to the proposed normal approximation. Combining \eqref{eq:diaz-francesNappr} with the linear transformation in Thm. \ref{thm:pham-gia_pdf_T}  between $T$ and $W$, we can obtain and theoretically confirm the specific normal distribution approximation of Hake's empirical distribution based on real data \citep{coletta_evidence_2023}.

\section{Numerical calculations for the ratio $X_1/X_2$}\label{sec:3_calculations}

\subsection{Mellin transform approach via convolution}\label{sec:3.1 Mellin transform}

As mentioned in the introduction, the only available specialized software for the ratio of normal RVs that we identified is the 20-year-old $C$ code from \cite{marsaglia_ratios_2006}. Additionally, we found an open-source solution capable of handling probabilistic and statistical computations for ratio distributions. Specifically, the more general Python package PaCAL (the Probabilistic CALculator; \cite{korzen_pacal_2014}) allows numerical arithmetic operations ($+, -, *, /$) and functions ($\exp, \log, \sqrt, \, \ldots)$  on RVs and runs on Cython (see Appendix \ref{app:tools}). 

The package primarily handles independent RVs (continuous or discrete) with known analytic forms of their PDFs but also offers some functionality for two-variable cases of dependent RVs. PaCAL algorithms implement ratio distribution computations using Mellin transform theory, applying Clenshaw-Curtis (CC) quadrature with barycentric interpolation via Chebyshev polynomials \citep{jaroszewicz_arithmetic_2012}. 

The Mellin transform, defined as \citep{springer_algebra_1979}

\nvertspace \begin{equation}
	\mathcal{M}_s\{f(x)\} \equiv \int_0^{\infty} x^{s-1} f(x) \, dx,
	\label{eq:mellin_transform}
\end{equation}

\noindent represents a powerful tool for studying the distribution of products, ratios, and, more generally, algebraic functions of independent RVs. Its significance is analogous to that of the Fourier integral transform, which is connected to the characteristic function effectively used for distributions of sums \citep{springer_algebra_1979}. 

\vertspace
\begin{remark} \small
Arguably, the most comprehensive handbook on the topic of the Mellin transform is Springer’s book \citep{springer_algebra_1979}. The most comprehensive tables of properties and closed expressions for several integral transforms, including the Mellin transform, can be found in the Caltech Project, led by editor A. Erdélyi \citep{erdelyi_tables_1954,erdelyi_tables_1954-1}.
\end{remark}

\vertspace

It is worth noting that the Mellin transform exists for a real function $f(x)$ that is defined and single-valued almost everywhere for $x \geqslant 0$ and is absolutely integrable over the range $(0, \infty)$. The Mellin transform can be extended to the range $(-\infty, \infty)$. The importance of the Mellin transform in studying the product and ratio of independent random variables lies in the fact that they can be expressed using the so-called Mellin convolution \citep{springer_algebra_1979}.

\vertspace

\begin{theorem}[Mellin convolution for product and ratio] \Newline
	Let $X_1$ and $X_2$ be independent continuous RVs with PDFs $f_1(x)$ and $f_2(x)$. Then, the PDFs of the product and ratio $X_1 X_2$ and $X_1 / X_2$ are given by
	\nvertspace \begin{equation}
		(f_1 \odot f_2)(t) = \int_{-\infty}^{+\infty} f_1(x) f_2\left(\frac{t}{x}\right) \frac{1}{|x|} \, dx, \,
		(f_1 \oslash f_2)(t) = \int_{-\infty}^{+\infty} f_1(xt) f_2(x) |x| \, dx. 
		\label{eq:M_convolution}
	\end{equation}
\end{theorem}

\vspace{-6pt}

\noindent For the ratio $T = X_1 / X_2$ from Thm.~\ref{thm:pham-gia_pdf_T}, where independent random variables $X_1 \sim \mathcal{N}(a, 1)$ and $X_2 \sim \mathcal{N}(b, 1)$, substituting the PDFs of $X_1$ and $X_2$,
  \nvertspace \begin{equation*}
f_1(x) = (1/\sqrt{2\pi}) \exp(-(x-a)^2/2),    
f_2(x) = (1/\sqrt{2\pi}) \exp(-(x-b)^2/2)   
   \end{equation*} 

\vspace{-6pt}

\noindent into the Mellin convolution \eqref{eq:M_convolution} gives us the PDF of the ratio distribution
\nvertspace \begin{equation}
	f_T(t) \equiv (f_1 \oslash f_2)(t) = \frac{1}{2\pi} \int_{-\infty}^{\infty} \exp\left(-\frac{(xt - a)^2 + (x - b)^2}{2}\right) |x| \, dx.
	\label{eq:T_convolution}
\end{equation}

\vspace{-6pt}

At first glance, this integral appears significantly simpler compared to expression \eqref{eq:f_T_pham-gia} in Thm. \ref{thm:pham-gia_pdf_T}, which contains Kummer’s confluent hypergeometric function. However, by using tabulated expressions of the Mellin transform \citep{erdelyi_tables_1954}, the Mellin convolution integral \eqref{eq:T_convolution} can be directly computed analytically to yield a result identical to \eqref{eq:f_T_pham-gia}; see Appendix \ref{app:proofs} (\textit{Proof of the analytic form of the integral} \eqref{eq:T_convolution}).

In practice, however, the integral 	\eqref{eq:T_convolution}, or more generally integral \eqref{eq:M_convolution} for the ratio of any (non)normal RVs,  can be computed numerically using an appropriate numerical method, quadrature. The Cython PACAL package calculates such integrals using the CC quadrature with barycentric interpolation.

\subsection{Fourier transform approach via characteristic function}

The probability distribution of any real-valued RV is completely determined by its characteristic function $\varphi_X(t) = \E{e^{itX}}$, which is mathematically equivalent to the Fourier transform of the probability measure of the RV \citep[chap. 3]{severini_elements_2011}. Unlike the PDF, the characteristic function (CF) of a random variable always exists, and in many situations, the form of the characteristic function is analytically known or much simpler than the PDF or CDF of the RV. 

In such cases, the characteristic function provides an alternative numerical method for effectively computing the PDF and CDF of the RV through the conceptually straightforward approach known as the \textit{numerical inversion of the characteristic function}, which remains less widely known but has been periodically highlighted in research over the past 50 years (e.g., \cite{davies_numerical_1973, waller_obtaining_1995, witkovsky_numerical_2016}). This computational approach is based on the numerical quadrature of the Gil-Pelaez inversion formulae \citep{gil-pelaez_note_1951}
\nvertspace \begin{equation}
	f(x) = \frac{1}{\pi} \int_0^\infty \Re \left[e^{-itx}\varphi(t)\right] \, dt, \quad F(x) = \frac{1}{2} - \frac{1}{\pi} \int_0^\infty \Im \left[\frac{e^{-itx}\varphi(t)}{t}\right] \, dt. \label{eq:numinvforms}
\end{equation}

\vspace{-15pt}

\subsubsection*{The current numerical inversion CF approach}

Regarding ratios of RVs, the currently used characteristic function approach \citep{witkovsky_charfuntool_2023} is based on the logarithmic transformation $Y = X_{1}^{c_{1}} \times \cdots \times X_{n}^{c_{n}}$ and applies only to $X_{j}$, which are assumed to be independent non-negative RVs. For $n=2$, $c_1=1$, $c_2=-1$, and assuming we know the CFs of the log-transformed variables, the CF of the ratio is given by $\varphi_{\log Y}(t) = \varphi_{\log X_1}(t) \times \varphi_{\log X_2}(-t)$. Inverting the CF of the log-transformed ratio $Y$, we can obtain the required distribution of the ratio $Y$. This approach can be successfully applied to model data with skewed distributions or distributions with heavy tails, where it is possible to compute exact distributions for likelihood ratio tests (LRT) statistics  \citep{witkovsky_logarithmic_2015} or, in more complex cases, in multivariate analysis involving LRT statistics \citep{filipiak_testing_2023}.

\vertspace

\begin{remark}\label{rem:divergence} \small It's worth noting that, for certain families of RVs, LRT statistics can be derived without relying on CFs or PDFs. In the gamma model, the Kullback–Leibler divergence can be decomposed into two independent components, whose exact distributions can be obtained through geometric integration and convolution approaches \citep{stehlik2014}. Such special-case derivations can serve as important cross-checks for numerical methods.
\end{remark}

\subsubsection*{An approach via Mellin convolution with CFs} 
In the case of independent RVs with values across the real line, a different approach is required. One straightforward option is to use the Mellin convolution \eqref{eq:M_convolution} and substitute for PDFs of $X_1$ and $X_2$ using the inversion formula \eqref{eq:numinvforms}
\nvertspace
\begin{equation} 
	f_T(t) = \frac{1}{\pi^2} \int_{-\infty}^{+\infty} \int_0^{\infty} \int_0^{\infty} \Re\left[e^{-i v x t} \varphi_{X_1}(v)\right] \Re\left[e^{-i u x} \varphi_{X_2}(u)\right] |x| \, dv \, du \, dx, \label{eq:MellinCF}
\end{equation}
\vspace{-6pt}

\noindent where for the $T$ ratio from Thm.~\ref{thm:pham-gia_pdf_T} CFs $ \varphi_{X_1}(v) = e^{i a v - \frac{v^2}{2}}$ and $ \varphi_{X_2}(u) = e^{i bu - \frac{u^2}{2}}$.

\subsubsection*{The Broda-Kan $X_1-wX_2$ approach}

\cite{broda_distributions_2016} proposed a different approach. They introduced an auxiliary random variable \( U = X_1 - wX_2, \, |w| < \infty \), which "linearizes" the problem of the CDF of the ratio in the sense that
\nvertspace \begin{equation*}
W = X_1\big/X_2 < w \Leftrightarrow (X_2 > 0, X_1 - wX_2 < 0) \text{ or } (X_2 < 0, X_1 - wX_2 > 0).
\end{equation*}

\vspace{-6pt}

\noindent Using an elementary identity \citep[see, e.g.,][]{renyi_probability_2007} for the probability of a symmetric difference $ \pr(A \Delta B ) = \pr(A) + \pr(B) - 2\pr(A \cap B) $  applied to the events \linebreak $(U < 0), (X_2 < 0)$, the CDF of the ratio is given by
\nvertspace
\begin{align*}
	F_W(w) &= \pr(U > 0, X_2 < 0) + \pr(U < 0, X_2 > 0) = \pr(U < 0 \, \Delta \, X_2 < 0) \\
	&= \pr(U < 0) + \pr(X_2 < 0) - 2 \pr(U < 0, X_2 < 0).
\end{align*}
Moreover, acknowledging the CF relation $\varphi_{U, X_2}(s, t) = \varphi_{X_1, X_2}(s, t - w s)$, the authors derive a general inversion theorem for the CDF and PDF of the ratio of not necessarily independent variables, as rigorously stated in Appendix \ref{app:proofs} (Thm.~\ref{thm:FR_Broda}). Here we will focus on the expression for the PDF

\vspace{-6pt}

\nvertspace \begin{equation}
	f_W(w) = \frac{1}{\pi^2} \int_{0}^{\infty} \int_{-\infty}^{\infty} \Re \left[ \frac{1}{t} \left. \frac{\partial \varphi_{X_1, X_2}(s,\tau)}{\partial \tau}  \right|_{\tau=-t - ws} \right] \, ds \, dt.
	\label{eq:B_num_inv_gen_formula}
\end{equation}
Considering $T = X_1\big/X_2 $ for independent $ X_1,X_2$, i.e., $ \varphi_{X_1, X_2}(s, t) = \varphi_{X_1}(s) \cdot \varphi_{X_2}(t) $, the PDF of $ T $ ratio takes the simplified form
\nvertspace \begin{equation}
	f_T(x) = \frac{1}{\pi^2} \int_{0}^{\infty} \int_{-\infty}^{\infty} \Re \left[ \frac{\varphi_{X_1}(s)}{t} \,  \varphi_{X_2}'(\tau)\Big|_{\tau=-t - xs} \right] \, ds \, dt.
	\label{eq:B_num_inv_formula}
\end{equation}
For the $T$ ratio from Thm.~\ref{thm:pham-gia_pdf_T}  we have $\varphi_{X_2}'( \tau) = \left( i b - \tau  \right) e^{i  b\tau - \frac{\tau^2}{2}}$.

\vertspace

\begin{remark}\label{rem:DEquad} \small
Similarly, as in the case of the convolution integral \eqref{eq:T_convolution}, it is naturally expected that the direct analytical integration of the particular forms of integrals \eqref{eq:MellinCF}, \eqref{eq:B_num_inv_formula} for the $T$ ratio of normal RVs must lead to the closed-form \eqref{eq:f_T_pham-gia} or with \eqref{eq:FvsERF}, as described in Thm.~\ref{thm:pham-gia_pdf_T}. Moreover, our considerations also point out that an iterated analytical integration of the general form \eqref{eq:MellinCF} should lead to the general Broda-Kan form \eqref{eq:B_num_inv_formula}. However, only general numerical quadratures and corresponding implemented algorithms for computing ratio distributions with various (non)normal RVs fulfill our goal.  
\end{remark}

\subsection{Numerical quadratures for the  integral transforms}

On the basis of our research and theoretical considerations, we have selected two fundamentally suitable approaches to compute the distribution of the ratio  of two independent RVs, $X_1$ and $X_2$, numerically. Both approaches are applicable under very mild conditions, i.e., without the necessity of assuming the normality of the RVs. 

The first approach is \textit{the 1D Mellin convolution integral} \eqref{eq:M_convolution} using the PDFs of $X_1$ and $X_2$. The second approach relies on knowing the CFs of $X_1$ and $X_2$, involving the calculation of\textit{ the 2D numerical CF inversion integral} \eqref{eq:B_num_inv_formula}. 

From a computational perspective, both integral transforms can be evaluated using any suitable numerical quadrature method. The open-source package PaCAL \citep{korzen_pacal_2014} employs the very effective CC quadrature with barycentric interpolation based on Chebyshev polynomials for the numerical calculation of \eqref{eq:M_convolution} \citep{jaroszewicz_arithmetic_2012}.

We also found that 2D numerical CF inversion integrals can be computed numerically using free MTB’s CharFunTool by \cite{witkovsky_charfuntool_2023} with the bivariate algorithm \texttt{cf2Dist2D} \citep[see theoretical details and implementation steps in][]{mijanovic_numerical_2023}. This algorithm also employs CC quadrature with barycentric interpolation in 2D and additionally offers a simple 2D trapezoidal rule (TR) quadrature as an alternative.

It is important to note that the output of \texttt{cf2Dist2D} is the PDF of the bivariate distribution of the random vector $(X_1, X_2)$, calculated from its bivariate CF $\phi$ by
\nvertspace \begin{equation}
	f(x_1, x_2) = \frac{1}{2 \pi^2} \int_0^{+\infty} \int_{-\infty}^{+\infty} \Re\left[ e^{-i\left(t_1 x_1 + t_2 x_2\right)} \phi(t_1, t_2) \right] \, dt_1 \, dt_2.
	\label{eq:joint_pdf}
\end{equation}
To compute our 2D integral \eqref{eq:B_num_inv_formula}, it is sufficient to choose \( f_T(x) = f(x,1) \), with the bivariate CF
$\phi(t_1,t_2) = 2 e^{i(x_1t_1+x_2t_2)} \varphi_{X_1}(t_1) \phi_2(-x_1t_1 - x_2t_2)/t_2; \phi_2(\cdot)  \equiv \varphi_{X_2}'(\cdot)$. The derivative of $\varphi_{X_2}$ can be calculated symbolically or numerically.

Regarding numerical quadrature, a very promising alternative to the CC algorithm is the double exponential (DE) quadrature \citep{takahasi_double_1974, mori_discovery_2005}. We recall only the core idea (for more details, see Remark \ref{rem:DEquad}), which lies in applying the simple trapezoidal rule to an integral transformed by a suitable DE transformation ($x=\Psi(t); \alpha, \beta \in \mathcal {R} \cup \left\{-\infty ,+\infty \right\}$):
\nvertspace \begin{equation*}
\int_\alpha^{\beta } f(x) \, dx \underset{x\,=\, \Psi(t)}{=} \int_{-\infty}^{\infty} f\Big(\Psi(t)\Big) \Psi'(t) \, dt \approx h \sum_{k=-n}^{n} f\Big(\Psi(k h)\Big) w(k h), \\[-6pt]
\end{equation*}
so the trapezoidal rule is modified by introducing a weight function $w = \Psi'(t)$ with a double-exponential decay rate
$
\left| \Psi'(t) \right| \underset{t \rightarrow \pm\infty}{\rightarrow} \exp\left(-\lambda\cdot \exp(\gamma |t|)\right);\lambda, \gamma \in \mathcal {R}^+.
$
Such DE transformations (three examples are in Table \ref{tab:integral_transforms}) make the trapezoidal rule optimally efficient, yielding exponentially fast and accurate results for a very broad class of functions (even with singularities).


\begin{table}[h!]
	\centering
	\caption{Examples of double exponential (DE) transformations for DE quadrature.}
	\begin{tabular}{c}
		\toprule
		DE transformation formulas \\ \midrule
		$\displaystyle\int_\alpha^\beta  f(x) \, dx, \quad x = \frac{\beta -\alpha}{2} \, \Psi(t) + \frac{\beta+\alpha}{2}, \quad \Psi(t) = \tanh\left(\frac{\pi}{2}  \sinh t\right)$ \\[12pt]
		$\displaystyle \int_\alpha^{\infty} f(x) \, dx, \quad x = \alpha + \Psi(t), \quad \Psi(t) = \exp\left(\frac{\pi}{2} \sinh t\right)$ \\ [12pt]
		$\displaystyle \int_{-\infty}^{\infty} f(x) \, dx, \quad x = \Psi(t), \quad \Psi(t) = \sinh\left(\frac{\pi}{2} \sinh t\right)$ \\
		\bottomrule
	\end{tabular}
	\label{tab:integral_transforms}
\end{table}

\vertspace

\begin{remark}\label{rem:DEquad} \small
	The DE quadrature was successfully applied in our study \citep[see DE quadrature refs there]{hancova_practical_2022} to calculate the gamma difference distribution with unequal shape parameters through numerical integration of the classical convolution integral or the 1D numerical CF inversion integral. DE quadrature formulas have been proven optimal, achieving smaller errors than any other formula with the same average node count \citep{sugihara_optimality_1997}. Although still less common in statistical applications, DE quadrature also has strong applications across fields such as molecular physics, fluid dynamics, boundary element methods, integral equations, ordinary differential equations, or complex indefinite and multiple integrals.
\end{remark}

\vertspace

In the Scientific Python ecosystem, the DE quadrature can be seamlessly combined with barycentric interpolation-based Chebyshev polynomials (via the \texttt{BarycentricInterpolator} function package \texttt{scipy}). For open-source computational implementations, the sinh-sinh DE transformation is particularly suitable for the 1D Mellin convolution integral \eqref{eq:M_convolution} in our case. This approach utilizes our Python implementation of the 1D DE quadrature \citep{gajdos_fdslrmgdd_2021}, which is a modified Python version of the original C implementation by \cite{ooura_oouras_2006} with  quadrature algorithm steps described in detail in \citet[Sec.5]{ooura_robust_1999}. 

Additionally, we have developed a small Python package, \texttt{Chebyshev.py}, specifically tailored for our PDF calculations, enabling the combination of DE quadrature with barycentric interpolation. Regarding the 2D numerical CF inversion \eqref{eq:B_num_inv_formula}, we adapted and adjusted Witkovský’s algorithmic approach for numerical inversion of the bivariate CF. Its full structure with implementation steps can be found in \citet[][Sec.2]{mijanovic_numerical_2023}. Key details of our modifications to numerically calculate the integral \eqref{eq:B_num_inv_formula} are provided in Appendix \ref{app:implementations} (\textit{Numerical 2D quadrature algorithm}).

Finally, for now, we eliminated the Monte Carlo approach, and the Mellin transform with CFs \eqref{eq:MellinCF} primarily due to their presumed very low computational speed. However, these approaches could provide benefits for future work, especially as valuable cross-checking methods in extended systematic numerical studies.

\section{A pilot numerical study}\label{sec:4_numerical_study}

\subsection{Setup, tools, and  conditions}

For purposes of numerical study benchmarking the chosen computational methods and digital tools, we use the open data science Jupyter interactive computing environment, combined with open-source tools like the computer algebra system SageMath as its kernel. This setup enable work with multiple programming languages, such as Python and R, within a unified digital workspace. Additionally, systems like SageMath provide the computational power of traditional software without financial barriers, fostering accessibility and empowering research without the constraints of commercial software. A detailed list of the used open digital tools is in Appendix \ref{app:tools}. 
The only commercial is MATLAB, but our code can run in the free alternative, Octave \citep{octave}.

All computations were conducted on a PC with Windows 11 and a 12th Gen Intel(R) Core(TM) i7-12700K processor @ 3.60 GHz, featuring Intel(R) UHD Graphics 770, 8 performance cores, 20 threads, and 32 GB of RAM.
Python, SageMath, and other free open software were installed from their official repositories, while officially licensed MATLAB (MTB) was run as a kernel in our Jupyter environment.

\subsubsection*{Goals, accuracy, and stability}

The main goal of the pilot study is to evaluate the described computational approaches in the case of a random variable $T$ from Thm. \ref{thm:pham-gia_pdf_T}. 
For each pair $(a,b)$, see details in Appendix \ref{app:implementations}, we conducted one numerical experiment for each considered computational method and its corresponding available digital tool. Each experiment was repeated in three runs, with each run containing at least 10 realizations of computing the PDF of $T$ at $N=1000$ points uniformly distributed over a chosen interval (see Appendix \ref{app:implementations}). This setup resulted in a total of $3 \times 10^4$ calculations per experiment. Execution time of each experiment was measured by built-in commands or functions.

Numerical accuracy was monitored via the maximum absolute error \\[-6pt]
$$\varepsilon_{\max} = \max_{i=1, \ldots, N} \left|\hat{f}_T(x_i)-f_{T\text{ref}}(x_i)\right|,$$  
where $\hat{f}_T(x_i)$ represents the values generated by a given computational approach. The reference values $f_{T\text{ref}}(x_i)$ were taken from \eqref{eq:f_T_pham-gia} using \eqref{eq:FvsERF}, computed using SageMath with quadruple precision (34-digit accuracy) and cross-checked against the arbitrary-precision C library Arb. The pilot study uses double precision (15-digit accuracy).

The predetermined number of repeated runs and realizations was selected to monitor the stability of the computational algorithms. In the case of runtime stability, we monitored the runtime coefficient of variation. As for numerical stability, error control was tracked by observing the order of the maximal error across repeated runs.

\subsection{Results and discussion}

The complete numerical study encompasses  1000 systematically carried-out experiments, 250 experiments for each pair $(a,b)$, where the benchmark for specific methods and tools consists of

\begin{itemize}
	\setlength\itemsep{3pt}
	\item \textit{13 experiments} using the analytic expression \eqref{eq:f_T_pham-gia} in SageMath via the \texttt{fast\_float} routine (53-bit, $\epsr = 10^{-3}, 10^{-4}, \ldots, 10^{-15}$);
	\item \textit{26 experiments} using CC quadrature of the 1D Mellin convolution integral \eqref{eq:M_convolution} in PaCAL, with and without Cython initialization (53-bit, $\epsr = 10^{-3}$ to $10^{-15}$);
	\item \textit{78 experiments} using DE quadrature of the 1D Mellin convolution integral \eqref{eq:M_convolution} in Python, with and without Numba, parallelization, barycentric interpolation (53-bit, $\epsr = 10^{-3}$ to $10^{-15}$); 
	\item \textit{15 experiments} using trapezoidal rule (TR) of the 2D numerical CF inversion integral \eqref{eq:B_num_inv_formula} in Python (53-bit) dealing with different algorithms for an approximating integral sum (compiled in different Numba settings -- with/without vectorization, parallelization, symbolic/numerical derivative); 
	\item \textit{118 experiments} using TR or CC quadrature (without/without vectorization, parallelization,  symbolic/numerical derivative, and various numbers of Chebyshev nodes; using built-in integrators and/or analytical implementations) of the 2D numerical CF inversion integral \eqref{eq:B_num_inv_formula} in MTB (53-bit).
\end{itemize}

\vertspace

\begin{figure}[h!]
	\centering
	\includegraphics[width=0.75\textwidth]{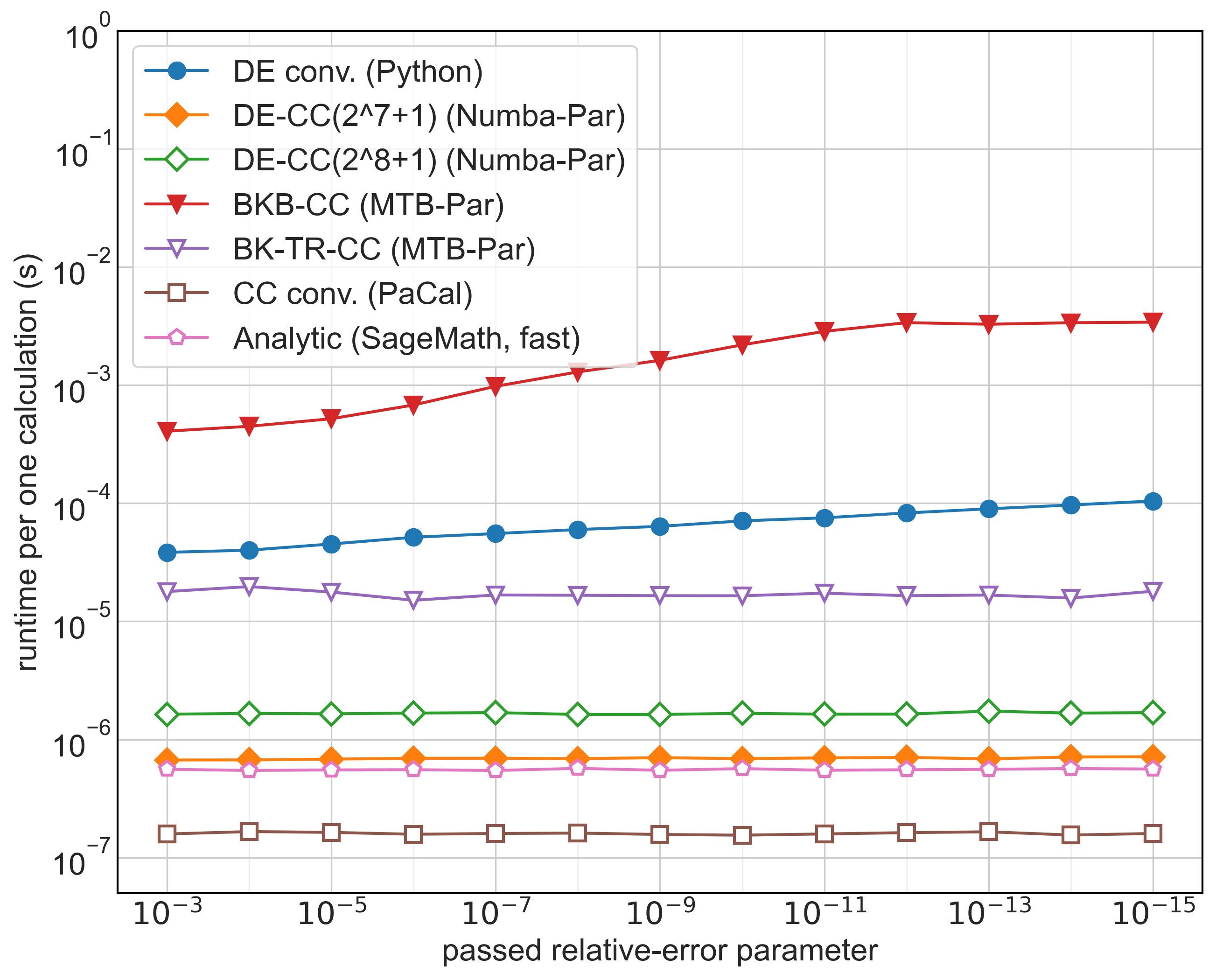}
	\caption{CPU runtimes of chosen methods and tools}
	\label{fig:runtimes_plot}
	\centering
	\includegraphics[width=0.75\textwidth]{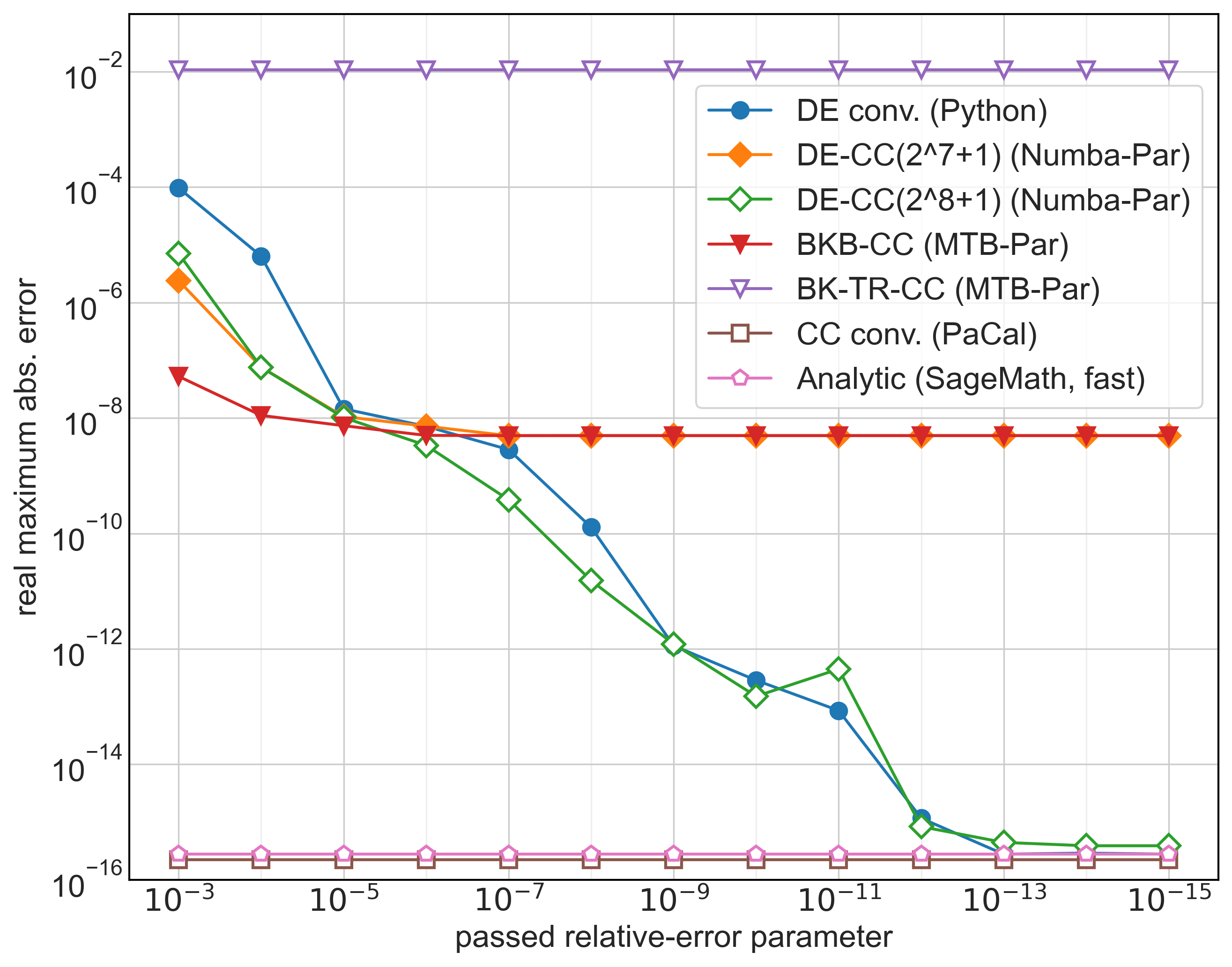}
	\caption{Real absolute errors of chosen methods and tools }
	\label{fig:error_plot}
\end{figure}

Each experiment benchmark results ($3 \times 10^4$ calculations) include average runtime, runtime standard deviation, acceleration (relative to the standard Python DE implementation), and accuracy (maximum absolute error). In Tab. 2, we present a sample of deliberately selected experiments in the case of $(a,b)=(1.5,1)$ divided into three main groups according to the computational method for the $T$ ratio distribution. Each group clearly demonstrates the key results of our numerical study. Some of these results are also visualized in Fig. \ref{fig:runtimes_plot} and \ref{fig:error_plot}.

\vspace{-12pt}

\begin{table}[h!]
	\renewcommand{\arraystretch}{1.25}
	\captionsetup{font=small}
	\caption{A sample of average runtimes, speedups, and real accuracy (maximum absolute error) for a ratio PDF, with each row summarizing one experiment.}
	\label{tab:numstudy}
	\footnotesize
	\begin{tabular}{clccc}
		\toprule
		\multicolumn{5}{c}{\small Probability density function $f_T(x)$ for ratio ($1000$ points over the chosen interval)} \\[3pt] 
		\midrule
		Method & Quadrature, Tool & Run~(s) & Accel & Accur\\[3pt] 
		\midrule
		\multirow{2}{*}{\footnotesize$\left[\begin{array}{c} \text{analytic} \\ \eqref{eq:f_T_pham-gia}\end{array} \right]$}
		& Kummer's ${}_1F_1$, Sage (\verb|fast_float|, 53-bit) & $5.56$e-04 & $69.0$ & $3$e-16 \\
		& Kummer's ${}_1F_1$, MTB (53-bit)                  & $5.88$e-03 & $6.53$ & $4$e-16 \\[3pt]
		
		\midrule
		\multirow{7}{*}{\footnotesize$\left[\begin{array}{c}\text{1D Mellin} \\ \text{convolution}\\ \text{integral \eqref{eq:M_convolution}}  \end{array} \right]$}
		& CC, PACAL (Cython, 53-bit)                & $1.60$e-04   & 240 & $2$e-16 \\
		
		\cmidrule{2-5}
		
		& DE, Python ($\epsr=1$e-03)         &  $3.84$e-02  & $1.00$  & $1$e-04 \\
		& DE, Python (Numba-par, $\epsr=1$e-15) &  $4.93$e-04  & $77.9$ & $3$e-16 \\
		& DE, Python (Numba-par, $\epsr=1$e-10) &  $3.67$e-04  & $105$ &  $3$e-13 \\
		& DE, Python (Numba-par, $\epsr=1$e-03)  &  $2.20$e-04  & $175$ & $1$e-04  \\
		& DE-CC($2^7+1$), (Numba, vect, $\epsr=1$e-06) &  $6.82$e-04  & $56.3$ & $7$e-09 \\
		& DE-CC($2^8+1$), (Numba, vect, $\epsr=1$e-15) &  $1.69$e-03  & $22.7$ & $4$e-16 \\
		
		\midrule

		\multirow{9}{*}{\footnotesize$\left[\begin{array}{c} \text{2D numerical} \\ \text{CF inversion} \\  \eqref{eq:B_num_inv_formula} \end{array} \right]$} 
		& TR, Python (Numba-par, no vect) & $1.20$e-01 & $0.320$ & $1$e-04  \\
		& TR, Python (Numba-par, vect) & $5.00$e-01 & $0.077 $ & $1$e-04  \\
		& BI, MTB (par, $\epsr=1$e-03) & $2.45$e-00 & $0.016$ & $1$e-08  \\
		& BI-CC, MTB (par, $\epsr=1$e-03) & $1.15$e-00 & $0.033$ & $5$e-08  \\
		& TR, MTB (par, no vect) & $7.97$e-00 & $0.005$ & $1$e-04  \\
		& TR, MTB (no par, vect) & $1.86$e-01 & $0.206$ & $1$e-04  \\
		& CC$(2^5+1)$, MTB (no par, vect) & $9.95$e-03 & $3.86$ & $1$e-02  \\
		& CC$(2^8+1)$, MTB (no par, vect) & $4.71$e-02 & $2.58$ & $1$e-02  \\
		\bottomrule
		\multicolumn{5}{l}{$^{\phantom{*}*}$\footnotesize \textbf{Abbr.} par~$\equiv$ ~parallelization, {vect}~$\equiv$~vectorization, BI~$\equiv$~built-in integrator, MTB $\equiv$ MATLAB
	} \\
		\multicolumn{5}{l}{\phantom{xxxxxxx} $2^k+1 \equiv$ the number of Chebyshev's nodes in barycentric interpolation}
		
	\end{tabular}
\end{table}

\vspace{-12pt}

All resulting data and code, including 20 Jupyter notebooks, also
describing our developments of algorithms in Python and MTB, are available at our GitHub repository 
\citep{hanc_jupyterperhakeratio_2024}.

As for the analytic formula \eqref{eq:f_T_pham-gia}, the built-in implementations of Kummer’s confluent hypergeometric function for computing the PDF of the $T$ ratio are highly accurate and reliable, which is not always the case for hypergeometric functions \citep{johansson_computing_2019,hancova_practical_2022}. Notably, the SageMath version is an order of magnitude faster (about 7x) than MTB, despite MTB’s reputation for numerical efficiency.

For the Mellin convolution integral \eqref{eq:M_convolution}, successful experiments were conducted in Python using the PACAL package and Python-based DE implementations. Both CC and DE quadrature methods achieved comparable speeds, when combined with high-performance compilers (Cython, Numba).  While PACAL was the fastest, its runtime did not account for the approximately one-second-long initialization of the barycentric interpolant. Both quadratures provided reliable results, with DE offering efficient accuracy control—where increasing accuracy by 12 orders of magnitude resulted in a computational slowdown of less than one order of magnitude.

At first glance, our newly implemented DE quadrature combined with barycentric interpolation of adjustable accuracy appears several times slower. However, it is important to note that the built-in \texttt{scipy} implementation of the barycentric interpolator does not utilize process parallelization. Without parallelization, the speed of the DE quadrature alone is lower by one order of magnitude. Ultimately, the use of Chebyshev interpolation accelerated the computation while conserving result accuracy.

The majority of experiments were conducted using custom MTB and Python algorithms for the numerical inversion of the ratio distribution via the Broda-Khan approach \eqref{eq:B_num_inv_formula}. While this 2D method is more computationally intensive, achieving speeds two orders of magnitude lower than the 1D Mellin convolution, the accuracy and performance are still practically acceptable. At lower accuracy, performance is even comparable to the 1D Python's DE quadrature or MTB's analytical computations.

In Python, parallelizing three nested loops (TR, Numba-par) improved 2D inversion speeds by 5-7$\times$ per loop, proportional to the number of cores. Without parallelization, the speed drops by 3 orders of magnitude. In MTB, full vectorization (TR, par vect) reduced computation time significantly ($40\times$ -- from 7.97 s to 0.19 s), which proved to be much more efficient than MTB's built-in parallelization. With barycentric interpolation, MTB is fast comparably to its analytical ${}_1F_1$ computations.

Concerning the stability of the explored computational approaches, the given number of runs and realizations led to excellent runtime stability, as the runtime coefficient of variation never exceeded $5\%$.  As for numerical stability, we achieved a sufficient level since the maximum error maintained its order or, at most, changed by one order of magnitude. Regarding variations in parameters $(a,b)$, the results were very similar across all pairs, as shown in the case of $(a,b) = (1.5,1)$, indicating the robustness of the methods with respect to the shape characteristics of Hake's ratio distribution.

\section{Conclusions}

In the first part of the paper, we examined the principal statistical properties of Hake's normalized gain, a ratio of two correlated normal RVs. In the PER community, where the measure was introduced as an indicator of educational effectiveness, it has primarily been used empirically. The Hake ratio lacks moments (mean, variance) and can exhibit heavy tails. However, in practice, it can often be approximated by a normal distribution. Previous efforts to analyze its statistical properties, such as those by \cite{burkholder_examination_2020,coletta_evidence_2023}, have remained empirical, with minimal reflection on existing statistical literature. To address this, we have provided six interactive Jupyter SageMath notebooks at our GitHub \citep{hanc_jupyterperhakeratio_2024} that allow users to explore the statistical properties of Hake’s ratio through virtual experimentation. 

From a statistical standpoint, this first part reviews the most relevant studies on the statistical properties of ratios of normal random variables (RVs), covering analytical forms, qualitative shape characteristics, approximations, and available software.  The theoretical and empirical insights obtained in the first part have provided a foundation for our key original and most impactful contributions as presented in the second part. 

Specifically, we introduced two novel computational statistics approaches for PDF calculations of ratio distributions. Our first approach  employs, originally, one-dimensional DE quadrature to the Mellin convolution of the PDFs of ratio constituents, where we combined and implemented the DE sinh-sinh quadrature with barycentric interpolation—a pairing that, to our knowledge, has not been published elsewhere, and for which we also successfully realized a practical implementation using modern open data science tools.

The method's speed, accuracy, and reliability were tested in a pilot numerical case study (6 Jupyter notebooks with resulting data files) in the context of Hake’s normalized gain  as a ratio of two independent normal  RVs. We demonstrated that our high-performance Python implementation (with Numba) of DE quadrature for the Mellin ratio convolution achieved reliability and accuracy at least on par with the sophisticated Cython-based PaCAL package, with speeds comparable to those of the C programming language. Our approach, however, offers advantages in terms of easily adjustable error levels, allowing for speed adjustments up to three times .

The methods' speed, accuracy, and reliability were tested in a pilot numerical case study (6 Jupyter notebooks with resulting data files) in the context of Hake’s normalized gain, as a ratio of two independent normal RVs. We demonstrated that our high-performance Python implementation (with Numba) of DE quadrature for the Mellin ratio convolution achieved a level of reliability and accuracy at least on par with the sophisticated Cython-based PaCAL package, with speeds comparable to those of the C programming language. Our approach, however, offers advantages in terms of easily adjustable error levels, allowing for speed adjustments of up to three times compared to the default setting.

Our second proposed method is a two-dimensional Broda–Khan numerical CFs inversion, which provides broader applicability for computing ratio distributions in several aspects. It does not require knowledge of PDFs, as CFs alone are sufficient and can also be applied to RVs with negative values. In this case, our original contribution lies in developing a fully vectorized and computationally efficient implementation. We would like to stress that, to the best of our knowledge, no existing work or tool provides a functional and reproducible computational implementation.

Regarding the pilot study involving the Hake ratio, although this 2D method is more computationally intensive, it yields speeds approximately two orders of magnitude slower than the 1D Mellin convolution. However, its accuracy and performance remain practically acceptable. Its performance becomes comparable to the 1D DE quadrature or analytically based computations at reduced accuracy settings.

\begin{figure}[h!] \centering \includegraphics[width=\textwidth]{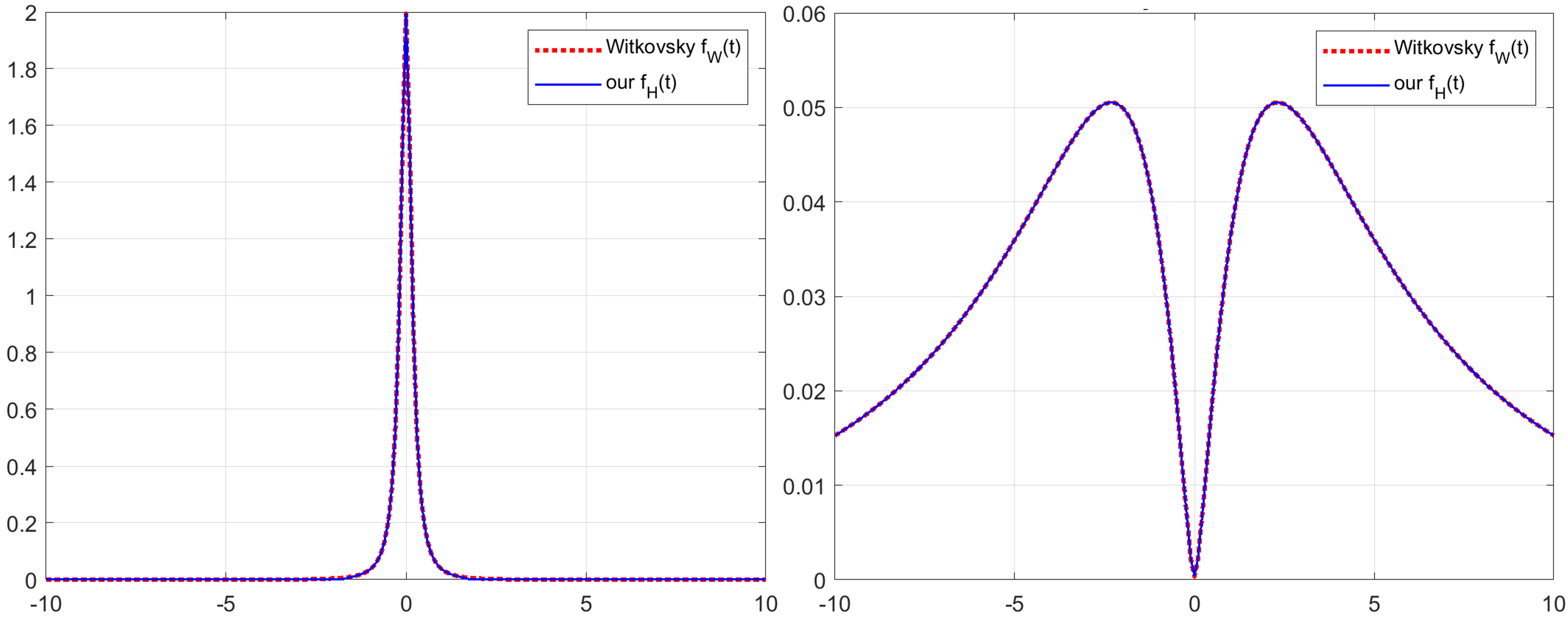} \caption{The ratio $X_1/X_2$ (left) and $X_2/X_1$ (right) for $X_1\sim N(0,1), X_2 \sim \chi^2(5)$.} \label{fig:other_examples} \end{figure}

Our general algorithmic implementation of the Broda–Khan inversion method for characteristic functions (CFs) also potentially extends beyond the assumption of normality of the random variables (RVs) considered in the ratio. To explore this capability, we have conducted preliminary calculations for ratios of various random variables, yielding promising results. For instance, in Fig.~\ref{fig:other_examples}, we compare the output of our method with an alternative PDF-based approach, demonstrating very good quantitative agreement. Furthermore, the method may also be applicable to correlated variables, further enhancing its versatility compared to the Mellin convolution approach.

In both approaches, we developed parallelized versions of algorithms, which are not available in existing computational tools. All developments and implementations were built using open modern data science technologies, making them widely accessible, transparent, and reproducible for the broader research community using statistics.

Finally, we acknowledge that our results hold only within the constraints of our pilot study, even though additional preliminary experiments were conducted, serving merely as indicators of emerging patterns for future research. To support more general conclusions, a much more extensive numerical study within the framework of a systematic design of experiments \citep[(DOE),][]{lorscheid_opening_2012} must be conducted.
Such a DOE study can identify key system variables, analyze their effects, optimize parameter settings for optimal responses, and enhance robustness against noise and uncontrollable factors. To effectively realize such studies, we see significant potential in higher-dimensional DE quadrature and GPU parallelization. Leveraging NVIDIA GPUs with 3000–4000 CUDA cores could accelerate computations by up to 1000 times. These research directions remain under active investigation.

\vspace{12pt}

\backmatter

\bmhead{Acknowledgments}
\small This work was supported by the Slovak Research and Development Agency under the Contract 
No. APVV-21-0369, No. APVV-21-0216 and by the Slovak Scientific Grant Agency VEGA under grant VEGA 1/0585/24. We sincerely thank the anonymous referees whose comments greatly enhanced the paper's clarity and consistency.

\vspace{12pt}

\begin{appendices}
	
\section{Hake's ratio interpretation}

Hake's ratio is widely used in the physics education community as a statistical measure of effect size and educational effectiveness, interpreting it as an indicator of the quality (effectiveness) of instruction or educational activities in introductory physics courses. Particularly, the ratio measures understanding of key physics concepts in mechanics and other fields through a specialized type of research-based standardized didactic test, known as concept inventories \citep{coletta_why_2020}.

\begin{figure}[h!]
	\centering
	\includegraphics[width=0.5\textwidth]{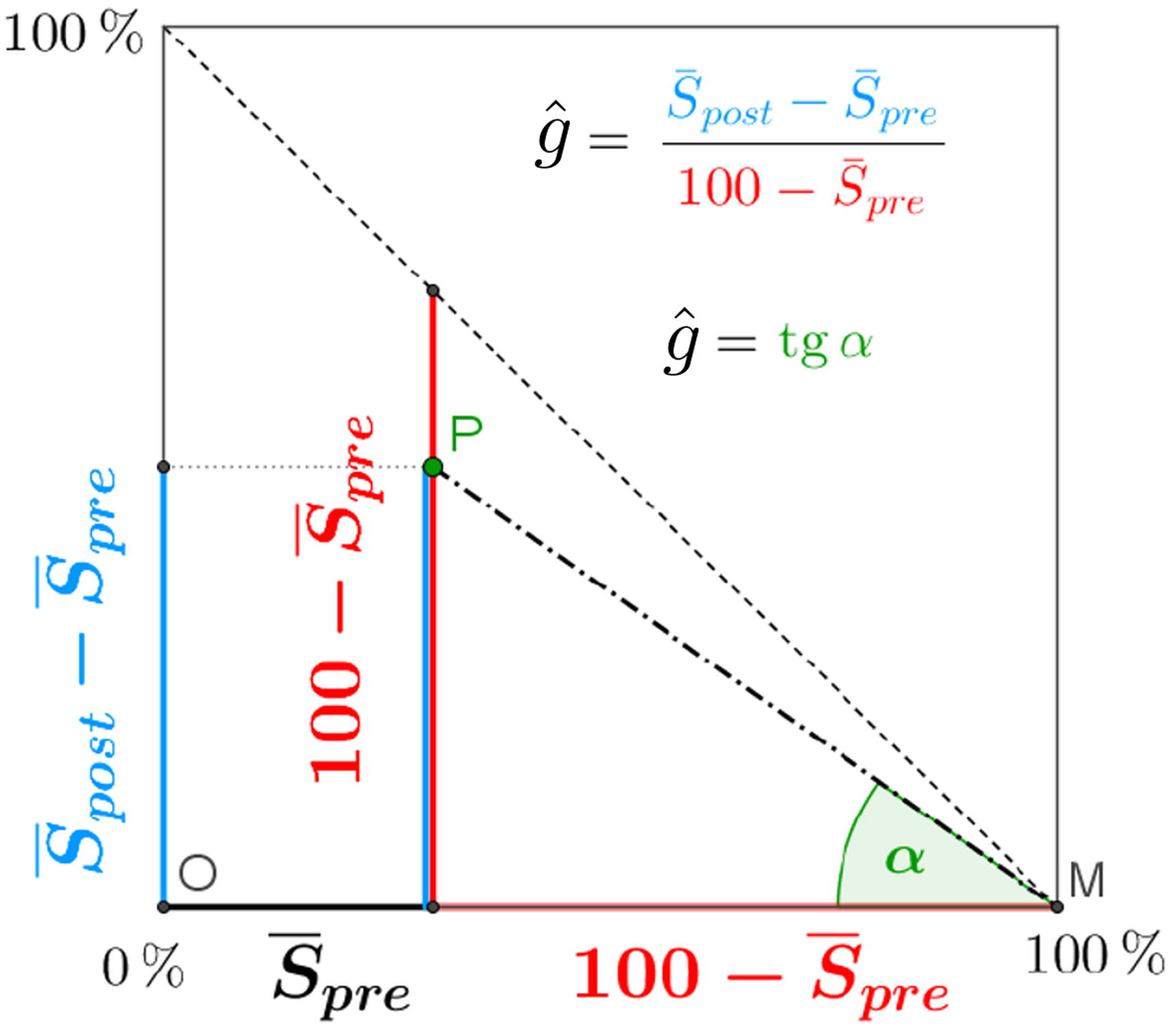}
	\caption{Geometric interpretation of Hake's ratio (adapted from \cite{strauch_measuring_2020}).}
	\label{fig:g_interpretation}
\end{figure}

When Hake introduced the normalized gain in his landmark meta-analysis article \citep{hake_interactive-engagement_1998} to measure instruction quality, he also used empirical data from 62 introductory physics courses ($N=6542$, average $n\approx 100$ participants per course). The data visualization provided him an important geometric interpretation with some consequences about the teaching methods used. At point $P$ (see Fig. \ref{fig:g_interpretation}), representing the results of a particular course, the Hake ratio is the tangent of the angle formed by the line connecting point $ P = (\Spre, \Spo - \Spre) $ with point $ M = (100, 0) $. It was observed that courses taught with traditional teaching methods achieved results up to $\Hg=0.3$, concentrated around a slope of $\Hg_{trad}=0.23$, while results of courses with interactive teaching methods fell within the range between lines with slopes from 0.3 to 0.6, centered around a slope of $\Hg_{int}=0.48$.

In education, Hake's statistic was and is understood almost exclusively as an empirical measure, representing the practical relative improvement of a group of students after instruction. Moreover, if a course attains an average normalized gain $\Hg$ below 0.3, it is considered low-effective; values of $\Hg$ between 0.3 and 0.7 are seen as medium-effective, and gains above 0.7 indicate high-effectiveness. Recurring discussions have emerged in the literature regarding the appropriateness and usefulness of this statistic \citep{savinainen_using_2002, hake_what_2015, coletta_why_2020}.


A common practice is to combine and interpret Hake's ratio with the popular and  standard measure of effect size (ES), known Cohen's $d$ and its sample-based estimator $\hat{d}$ \citep{cohen_statistical_1988,grissom_effect_2012} defined as

\begin{equation}
	d = \frac{\mpo - \mpre}{\s}, \qquad \hat{d} = \frac{\Spo - \Spre}{s}
	\label{eq:cohen_d} \\[12pt]
\end{equation}

\noindent where $\s, s$ are the population standard deviations  of the score and its  sample version.

\vertspace

\begin{remark} \small
	In definition~\eqref{eq:cohen_d}, we typically assume $\Di{\Sb}=\Di{\Sa}=\s^2$. There are various modifications of the formula depending on equal or unequal variances or types of considered random variables; see more in \cite{grissom_effect_2012}.
\end{remark}

\newpage

\section{Important formulas and proofs} \label{app:proofs}

\begin{proposition}[Hake's ratio distribution]\label{prop:g_distribution}
	Let $\Sb, \Sa$ be normal random variables, where $\Sa \sim \N(\mpre, \spre^2)$ and $\Sb \sim \N(\mpo, \spo^2)$ and $\Spre, \Spo$ are sample means for random samples of size $n$  from $\Sa, \Sb$.  Then Hake's normalized gain $\hat{g}$, given by \eqref{eq:normalized_gain} is a ratio $W$ of correlated normal random variables $X_1$ and $X_2$:
	\begin{equation}
		W = X_1 \big/ X_2; \quad X_1 = \Spo - \Spre, X_2 = 100 - \overline{S}_{\text{pre}}
		\label{eq:W_ratio}
	\end{equation}
	where 
	\begin{itemize}[leftmargin=10pt]
		\setlength{\itemsep}{6pt}
		\setlength{\leftmargin}{0pt}
		\item[] $(X_1, X_2) \sim \BN(\m_1, \m_2; \s_1, \s_2; \rho)$, \, $\m_1 = \mpo - \mpre$, \, $\m_2 = 100 - \mpre$,
		\item[] $n\s_1^2= \spre^2 + \spo^2 - 2 \spre\, \spo\, \rs$, \, $n\s_2^2 = \spre^2$, \, $\sqrt{n}\rho = \dfrac{\spre}{\s_1} - \rs \dfrac{\spo}{\s_1}$, \,  $\rs =\rba.$
	\end{itemize}
\end{proposition}

\noindent We have proven the relations between parameters as direct consequences of basic properties of covariance and correlation \citep{renyi_probability_2007}. 

\vertspace
\begin{remark} \small
	The strict normality assumption for scores $\Sa$ and $\Sb$ is not always necessary; asymptotic normality often suffices. By the Central Limit Theorem (CLT), sample means from non-normal distributions can be accurately approximated by a normal distribution if $n$ is sufficiently large—typically $n > 30$ \citep[p. 1563]{salkind_encyclopedia_2010}. However, for highly skewed data, stricter conditions may apply, requiring $n > 50$ or more \citep{lohr_sampling_2019}. In educational settings, considerably smaller sample sizes may be acceptable. For example, a 2018 Slovak national study (a representative sample of 40 secondary schools with 919 students) found that even $n \approx 15$ was sufficient in some cases \citep{strauch_measuring_2020}. However, careful considerations are still needed to determine whether the CLT is applicable.
\end{remark}

\vertspace

\noindent \textbi{Modality curve} (defined by $f'_T(t)=0$). 
According to \cite{marsaglia_ratios_2006}, the modality curve plotted in Fig.~\ref{fig:bimodal_distribution} may be numerically approximated by ($a_0 =2.256058904$):
\begin{equation*}
	b=\frac{18.621-63.411 a+84.041 a^2-54.668 a^3+17.716 a^4-2.2986 a^5}{a_0-a}, 1 \leq a<a_0 .
\end{equation*}
The coefficients were obtained using CAS Maple with 50 digits of accuracy.

\vertspace
\begin{corollary}[$a,b$ for Hake's ratio]\label{cor:Hake_ab} \Newline
	\nvertspace\begin{equation*}
		a = \frac{\rs}{\sqrt{1-\rho^{*2}}} \left( (100-\mpre)
		\left(\frac{1}{\spre}-\frac{1}{\spo\,\rs} \right) - \frac{\mpo-\mpre}{\spo\,\rs}\right)
	\end{equation*}
	\nvertspace\begin{equation*}
		b = 
		(100-\mpre)/\spre, \quad\rs =\rba
	\end{equation*}
\end{corollary}

\vertspace

\noindent\textbi{Proof of the analytic form of the integral \eqref{eq:T_convolution}} (via the Mellin transform). \\
 By changing to new variables $y$ and $q$ with the substitutions $x = \frac{\sqrt{2} y}{\sqrt{1 + t^2}}$ and $q = \frac{at + b}{\sqrt{t^2 + 1}}$, the Mellin convolution integral \eqref{eq:T_convolution} transforms to the integral
\begin{equation*}
	f_T(t)  = \frac{1}{\pi(1 + t^2)} e^{-\frac{1}{2}(a^2 + b^2)} \int_{-\infty}^{\infty} e^{-y^2 + \sqrt{2} q y} \cdot |y| \, dy 
\end{equation*}
\noindent Splitting the integral at zero and applying the Mellin transform, we obtain
\begin{equation*}
	f_T(t) = \frac{1}{\pi(1 + t^2)} e^{-\frac{1}{2}(a^2 + b^2)} 
	\left(\mathcal{M}_2\{e^{-y^2 + \sqrt{2} q y}\} + \mathcal{M}_2\{e^{-y^2 - \sqrt{2} q y}\}\right)
\end{equation*}
\noindent There is a closed integral expression for this Mellin transform \cite[eq. 6.3.(13),][]{erdelyi_tables_1954}:
\begin{equation*}
	\mathcal{M}_2\{e^{-y^2 \pm \sqrt{2} q y}\} = \frac{1}{2} \operatorname{exp}\left(\frac{q^2}{4}\right) \cdot D_{-2}[\mp q] = H_{-2}\left( \pm \frac{q}{\sqrt{2}}\right)
\end{equation*}
where $D_\nu$ is the parabolic cylinder function and $H_\nu$ is the Hermite function \citep{lebedev_special_1972}. Since the following relationship holds between the Hermite function $H_\nu$ and Kummer's confluent hypergeometric function $M(\alpha, \beta, z) \equiv {}_1 F_1\left(\begin{matrix} \alpha \\ \beta \end{matrix} ; z\right)$ \citep{gradshteyn_table_2007}
\begin{equation*}
	H_{-2}\left(\frac{q}{\sqrt{2}}\right) + H_{-2}\left(-\frac{q}{\sqrt{2}}\right) = {}_{1} F_{1}\left( \begin{matrix} 1 \\ 1/2 \end{matrix} ; \frac{q^2}{2} \right),
\end{equation*}
\noindent we obtain the desired result \eqref{eq:f_T_pham-gia}
\begin{equation*}
	f_T(t) = \frac{1}{\exp\left((a^2 + b^2)/2\right) \pi(1 + t^2)} \, {}_1F_1\left( \begin{matrix} 1 \\ 1/2 \end{matrix} ; \frac{(at + b)^2}{2(1 + t^2)}\right).
\end{equation*}

\vertspace

\begin{theorem}[The inversion theorem for a ratio distribution \citep{broda_distributions_2016}]\label{thm:FR_Broda} \Newline
	If $(X, Y)$ has a finite mean, $\varphi_{X, Y}$ is absolutely integrable, and 0 is not an atom of $X - r Y$ $(\text{i.e., }  P(X-rY=0)=0)$, then for $R= X/Y; \,|r| < \infty$ the CDF and PDF
	is
	 \begin{equation}
		F_R(r) = \frac{1}{2} + \frac{1}{\pi^2} \int_0^{\infty} \int_0^{\infty} \Re\left[\varphi_{X, Y}(s, t - r s) - \varphi_{X, Y}(s, -t - r s)\right] \frac{\mathrm{d} s}{s} \frac{\mathrm{d} t}{t}
	\end{equation}
	and
\begin{equation}
		f_R(r) = \frac{1}{\pi^2} \int_0^{\infty} \int_{-\infty}^{\infty} \Re\left[\varphi_{2}(s, -t - r s)\right] \mathrm{d} s \frac{\mathrm{d} t}{t}, \quad \varphi_2(s, t) \equiv \partial / \partial t \,\varphi_{X, Y}(s, t)
	\end{equation}
	whenever this integral converges absolutely.
\end{theorem}



\section{Computational implementations}\label{app:implementations}

\noindent\textbi{Numerical 2D quadrature algorithm for integral \eqref{eq:B_num_inv_formula}}. \\
The integral \eqref{eq:B_num_inv_formula} can be rewritten in the form corresponding to the 2D integral \eqref{eq:joint_pdf}
\begin{equation*}
	f_T(x) \equiv f(x,1) =  \frac{1}{\pi^2} \int_{0}^{\infty} \int_{-\infty}^{\infty} \Re \left[\phi_2\big(-(xt_1 + t_2)\big)\,  \varphi_{1d}(t_1,t_2)  \right] \, dt_1 \, dt_2
	\label{eq:B_num_inv_formula2}
\end{equation*}
where $\varphi_{1d}(t_1,t_2) \equiv \varphi_{X_1}(t_1)/t_2 $ and $\phi_2(\cdot) \equiv \varphi'_{X_2}(\cdot)$. 

\noindent Then, the 2D integral can be approximated by the trapezoidal rule integral sum \citep{mijanovic_numerical_2023}, giving the PDF at a point $x \in \mathcal{R}$ (for which the PDF exists) as
	$$
f_T(x) = f\left(x, 1\right) \approx \frac{1}{\pi^2} h_1 h_2 \sum_{\nu_1=-N}^{N} \sum_{\nu_2=0}^{N}  \Re\Big\{ \phi_2\big(- (x t_1 +  t_2)\big) \,\varphi_{1d}(t_1,t_2) \Big\}$$
\vspace{-6pt}
$$t_1=h_1\left(\nu_1+0.5\right), t_2=h_2\left(\nu_2+0.5\right)$$

\noindent 
Using vectorization and matrix representations as in \cite{mijanovic_numerical_2023}, we can get an
algorithm (see a schematic version Algorithm 1) for the ratio PDF \eqref{eq:B_num_inv_formula}, conceptually similar as for the bivariate PDF \eqref{eq:joint_pdf}.

\begin{algorithm*}
	\small
	\caption{Determination of the ratio PDF \eqref{eq:B_num_inv_formula} vs. bivariate PDF \eqref{eq:joint_pdf}}\label{algo1}
	\begin{algorithmic}[1]
		\State \texttt{x = [X1(:) X2(:)] \% all (x1,1)}    \hspace{\stretch{1}} \texttt{x = [X1(:) X2(:)] \% all (x1,x2)} 
		\State \texttt{t = [T1(:) T2(:)] \% all (t1,t2)}   \hspace{\stretch{1}} \texttt{t = [T1(:) T2(:)] \% all (t1,t2)}   
		\State \texttt{c = h(1)*h(2)/(pi*pi)} \hspace{\stretch{1}} \texttt{c = h(1)*h(2)/(2*pi*pi)\phantom{xxxxxxxx}} 
 		\State \texttt{\% ============== Integrand terms via vectorization ==============}
		\State \texttt{CF2\_dif = cf2\_dif(-x*t')} \hspace{\stretch{1}}  \texttt{E \phantom{xx}= exp(-i*x*t')\phantom{xxxxxxxxxxxxx}}
		\State \texttt{CF1\_d \phantom{x}  = cf1\_d(t)} \hspace{\stretch{1}}  \texttt{cft = cf(t)\phantom{xxxxxxxxxxxxxxxxxxxx}}
		\State \texttt{\% ============== Integral sum via vectorization ==============}
		\State \texttt{pdf = c*real(CF2\_dif*CF1\_d)}\hspace{\stretch{1}} \texttt{pdf = c*real(E*cft)\phantom{xxxxxxxxxxxx}}
	\end{algorithmic}
\end{algorithm*}

\vspace{12pt}

\noindent\textbi{Parameters (a,b) in PDF calculations}.\\
Since, during the pilot study, we decided to check results interactively and exploratorily, we limited numerical experiments to four pairs of $(a,b)$. A larger number of cases in this type of pilot study, aimed at gaining insights and understanding underlying patterns, would be highly time-consuming and practically infeasible.

Specifically, we considered $(a,b) = (2,0.25), (1.5,1), (4,7), (5,25)$, which were not chosen randomly. They represent four qualitatively distinct shape categories with respect to normality \eqref{eq:diaz-francesNappr}: (1) \textit{bimodal}, (2) \textit{unimodal with heavy tails} (bad normal approximation), (3) \textit{unimodal, skewed} (reasonable normal approximation), and (4) \textit{unimodal, symmetric} (good normal approximation). 	Qualitatively similar cases have been analyzed by \citet[their Tab.~1 and figs]{diaz-frances_existence_2013}.
Simultaneously, these values have practical meaning since they naturally emerged from our theoretical explorations or from real data in PER. For example, values $a \in (4,5), b\in(20,25)$ typically appeared in  our university physics courses \citep[years 2006–2020]{strauch_measuring_2020}. 

According to \citet[sec.~3, p.~4]{marsaglia_ratios_2006}, the heavy-tailed nature arises from the fact that the PDF \eqref{eq:f_T_pham-gia} is also a mixture of two distributions, one of which is a heavy-tailed Cauchy distribution that dominates when the factor $\exp(-\tfrac{1}{2}(a^2 + b^2))$ is not too small. It is also worth noting that, according to the rule of thumb and the modality curve, cases (3) and (4) are theoretically bimodal; however, the second (left) mode is extremely small ($f_T(\cdot) < 10^{-16}$), distant, and can be safely ignored \citep[sec.~3.1]{marsaglia_ratios_2006}.See details in Jupyter Notebook 2.1 – plots at GitHub, \cite{hanc_jupyterperhakeratio_2024} for our specific cases.

\vertspace

\noindent\textbi{Intervals and moments in PDF calculations}.\\
For Hake's ratio, the interval used for PDF calculations is determined by practical moments \eqref{eq:diaz-francesNappr} (even in cases of poor normal approximations): 
$\mu \pm k\,\sigma, \, \mu = a/b, \, \sigma^2 = \mu^2(1/a^2+1/b^2),$ 
where $k$ is a suitable integer. To keep simplicity, we set $k = 2$, providing a rough estimate of the confidence interval for the distribution. In the case $(a,b) = (1.5,1)$, whose results are presented in Sec. 4.2, the interval was approximately $(-5, 8)$. 

\vertspace
\begin{remark} \small
	In general, a more refined procedure is needed. According to \cite{mijanovic_numerical_2023}, estimates of the first and second moments of ratio constituents can be calculated from CFs as:
	$$
	\begin{aligned}
		& \hat{\mu}_i = \frac{8 \Im[\varphi_{Xi}(h)]}{5 h}-\frac{2 \Im[\varphi_{Xi}(2 h)]}{5 h}+\frac{8 \Im[\varphi_{Xi}(3 h)]}{105 h}-\frac{2 \Im[\varphi_{Xi}(4 h)]}{280 h}, \\
		& \hat{\mu}^*_i =\frac{205}{72 h^2}-\frac{16 \Re
			[\varphi_{Xi}(h)]}{5 h^2}+\frac{2 \Re[\varphi_{Xi}(2 h)]}{5 h^2}-\frac{16 \Re[\varphi_{Xi}(3 h)]}{315 h^2}+\frac{2 \Re[\varphi_{Xi}(4 h)]}{560 h^2}, 
	\end{aligned}
	$$
	where $h$ is a chosen tolerance for numerical differentiation $($default $h=10^{-4})$.
	Then, estimates of variances and coefficients of variation are $\hat{\sigma}_i^2 = \hat{\mu}^*_i - \hat{\mu}_i^2, \quad \hat{\delta}_i = \hat{\sigma}_i / \hat{\mu}_i.$ 
	More generally, \cite{witkovsky_numerical_2016} recommends a simple yet effective rule of thumb—the \textit{six sigma rule}, which, in the case of practical ratio moments, provides a rough interval:
	\nvertspace
	$$\hat{\mu}\pm k \,\hat{\sigma} = \hat{\mu}_1 /\hat{\mu}_2\pm k (\hat{\mu}_1 /\hat{\mu}_2)\sqrt{\hat{\delta}_1^2+\hat{\delta}_2^2}, \quad k = 6.$$ 
	This interval does not necessarily have to be symmetric, and $k_1, k_2$ can be chosen to meet a predefined criterion for neglecting sufficiently small numerical outputs below $\varepsilon=10^{-n}$.
	In the case where $\hat{\mu}_2 \approx 0$ (situation in Fig. \ref{fig:other_examples}), the interval can be calculated alternatively as a predefined-criterion extension of the union of six sigma intervals of the ratio constituents.
\end{remark}
\vertspace

\newpage

\section{Open digital tools} \label{app:tools}

\vspace{-30pt}

\begin{table}[h]
    \caption{Chosen digital tools (all open except MATLAB) for our numerical study}
	\begin{tabular}{@{}p{0.25\textwidth}p{0.75\textwidth}@{}}
		\toprule
		\parbox[t]{0.25\textwidth}{\raggedright \textbf{Main Tools}} & 
		\textbf{SageMath} \cite[v.~\!10.3,][]{sage}, \textbf{Python} \cite[v.~\!3.11.6,][]{python3}, \textbf{MATLAB} \cite[v.~\!R2023b,][]{matlab} \\
		\midrule
		\parbox[t]{0.25\textwidth}{\raggedright \textbf{Numerical \& scientific computing libraries}} & 
		\parbox[t]{0.75\textwidth}{\raggedright
			\begin{itemize}[leftmargin=*]
				\vspace{-6pt}
				\item \textbf{NumPy} \cite[v.~\!1.19.1,][]{numpy2020}: Numerical Python library for fast vector computations
				\item \textbf{SciPy} \cite[v.~\!1.11.3,][]{2020SciPy_NMeth}: Fundamental Python library for scientific computing
				\item \textbf{GSL} \cite[v.~\!2.7.1,][]{galassi_gnu_2009}: C, C++ numerical library for scientific computing
				\item \textbf{PaCAL} \cite[v.~1.6.1,][]{korzen_pacal_2014}: Cython probabilistic calculator for arithmetics of i.r.v.'s
				\item \textbf{Arb} \cite[v.~\!2.16.0,][]{johansson_arb_2014}: C library for arbitrary-precision interval arithmetic using the midpoint-radius representation ("ball arithmetic")
				\item \textbf{Jupyter-Matlab-Proxy} \cite[v.~\!0.15.3,][]{jupyter-matlab}: Python library from MathWorks enabling MTB as a kernel for Jupyter notebooks.
		\end{itemize}} \\
		\midrule
		\parbox[t]{0.25\textwidth}{\raggedright \textbf{High performance Python compilers}} &
		\parbox[t]{0.75\textwidth}{\raggedright
			\begin{itemize}[leftmargin=*]
					\vspace{-6pt}
				\item \textbf{Cython} \cite[v.~\!3.0.10,][]{cython}: Optimizing static compiler translating Python into C code
				\item \textbf{Numba} \cite[v.~\!0.59.1,][]{numba2015}: JIT compiler translating Python into fast machine code; also suitable for GPU/CPU parallelism
		\end{itemize}} \\
		\botrule
	\end{tabular}
\end{table}

\end{appendices}


\setlength{\bibsep}{3pt}
\small
\bibliography{references,software}

\end{document}